# High-Throughput Detection and Manipulation of Single Nitrogen-Vacancy Center's Charge in Nanodiamonds

*Maabur Sow[†], Horst Steuer[†], Sanmi Adekanye[‡], Laia Ginés[‖], Soumen Mandal[‖], Barak Gilboa[†], Oliver A. Williams[‖], Jason M. Smith[‡], Achillefs N. Kapanidis*[†]*

[†]Department of Physics, University of Oxford, Oxford OX1 3PU, UK

[‡]Department of Materials, University of Oxford, Parks Road, Oxford OX1 3PH, UK

[‖]School of Physics and Astronomy, Cardiff University, Cardiff CF24 3AA, UK

The fluorescent nitrogen-vacancy (NV) defect in diamond has remarkable photophysical properties, including high photostability, which allows stable fluorescence emission for hours; as a result, there has been much interest in using nanodiamonds (NDs) for quantum optics and biological imaging. Such applications have been limited by the complexity and heterogeneity of ND photophysics. Photophysics of the NV center in NDs have been studied before, but the lack of a sensitive and high-throughput method has limited the characterization of NDs. Here is reported a systematic analysis of NDs using two-color wide-field epifluorescence imaging coupled to high-throughput single-particle detection of single NVs in NDs with sizes down to 5-10 nm. By using fluorescence intensity ratios, the charge conversion of single NV center ($NV^-$ or $NV^0$) is observed and the lifetimes of different NV charge states in NDs is measured. The discovery of reversible manipulation of NV charge states by pH is also presented. In addition to provide another mechanism to control the NV charge state in nanodiamonds, our findings open the possibility to perform pH nanosensing with a non-photobleachable probe.

Nanodiamonds (NDs) as single-photon sources and bio-imaging probes have attracted significant interest in the past two decades.[1,2] A major reason for this attention is that, above a size of 35 nm, the fluorescent nitrogen-vacancy (NV) centers in the ND emit bright photoluminescence without blinking or photobleaching.[3] The NV crystal defect consists of a substitutional nitrogen atom adjacent to a carbon vacancy (Figure 1a) that can be incorporated inside NDs (5-200 nm) at different concentrations.[4] Moreover, NDs show little toxicity and are easy to functionalize.[5] These desirable properties have been exploited in *in vitro* single-molecule experiments and long (>10 min) intracellular tracking;[6-10] such tracking also enabled nanosensing inside living cells.[11,12]

A facile way for nanosensing is to detect changes in the charge of the NV center as charge transitions ($NV^-$ and $NV^0$) are triggered by chemical events or variations of electrical potential.[13,14] A more sensitive sensing approach exploits the spin-state-dependent fluorescence of the NV center, which can be manipulated at room temperature;[15-17] this property was harnessed to measure temperature changes inside living neurons.[18] *In vitro* experiments also demonstrated that NDs can detect down to a few atomic spin labels (*i.e.*, gadolinium atoms), whereas a single nitroxide label inside a protein was sensed using NV center in bulk diamond.[19,20]

Despite their promise, ND applications in bioimaging have been limited by the low brightness of sub 20-nm NDs, since a single NV center is 10-times less bright than a typical organic fluorophore. Moreover, it is still very difficult to manufacture small NDs suitable for high-sensitivity nanosensing (*i.e.*, magnetic field from individual molecules or atoms) as such NDs must contain few impurities (*e.g.*, nitrogen or $^{13}C$) and little crystal strain. Thus, there are numerous efforts to manufacture small, bright, and high-purity NDs that differ by size, nitrogen content and surface chemistry; these ND samples display different NV center emission spectra and intensity levels due to interactions with the surface, or due to a different number of NV centers per particle.[21, 22] For these reasons, the photophysical characterization of NDs is crucial to ensure successful applications in bioimaging, especially using single-molecule microscopy.

An important question in ND characterization is the proportion of NDs containing single NV centers, a property paramount for the optimization of ND manufacture and for ND applications as single-photon sources.[2, 23] The conventional method to confirm the presence of the single NV centers is to measure the coherence of its emission and calculate the probability of photons being emitted at the same time. However, such photon-correlation experiments require complex instrumentation and have limited throughput, since each measurement needs to be performed individually on NDs.[24, 25] An alternative method to identify single NV center is to measure the photon count corresponding to a single NV center; this, however, is also complicated by the orientation of the two NV center's orthogonal dipoles.[26] As a result, there is a need for a high-throughput method reporting on the fraction of single emitters in NDs.

Characterization methods are also essential for new ND bio-sensing assays. Petráková *et al.* showed that the charge state of a NV center could be used to detect chemical changes on the ND surface.[13, 27] Their approach is to measure changes in the NDs fluorescence (brightness or emission spectra) to differentiate the two photoactive charge states of the NV center ($NV^-$ and $NV^0$; $NV^+$ is non-photoactive), since $NV^0$ has its emission 60 nm blue-shifted compared to $NV^-$ (see Figure 1b).[13] The team then established that the charge state of NV centers in NDs is affected by specific chemical changes on the ND surface (*e.g.* modification of functional groups or adsorption of polymers). Nonetheless, the direct charge manipulation of single NV centers by pH in ND was not performed because the NDs used were too large (~49 nm).[27] Indeed (10-20 nm) NDs would be optimal for such experiments, since they contain shallower NV centers; unfortunately, working with small NDs poses technical challenges in terms of manufacture and detection capability (*i.e.* small NDs form aggregates and have low brightness).[4]

Here, we report a sensitive and high-throughput wide-field imaging approach that allows us to measure reliably single NV charge states and compare the proportion of single emitters in different ND samples (5-200 nm in size). Our approach provides hundreds of ND fluorescence time traces in seconds, which allows measurements of the proportion of NDs containing single NV centers, and the detection of spectral shifts that unravel changes in the NV center charge state. We use our method to study dynamic charge states transitions in multiple NDs and to demonstrate that the charge state in 10-nm NDs can be reversibly manipulated by pH, making NDs promising probes for pH sensing.

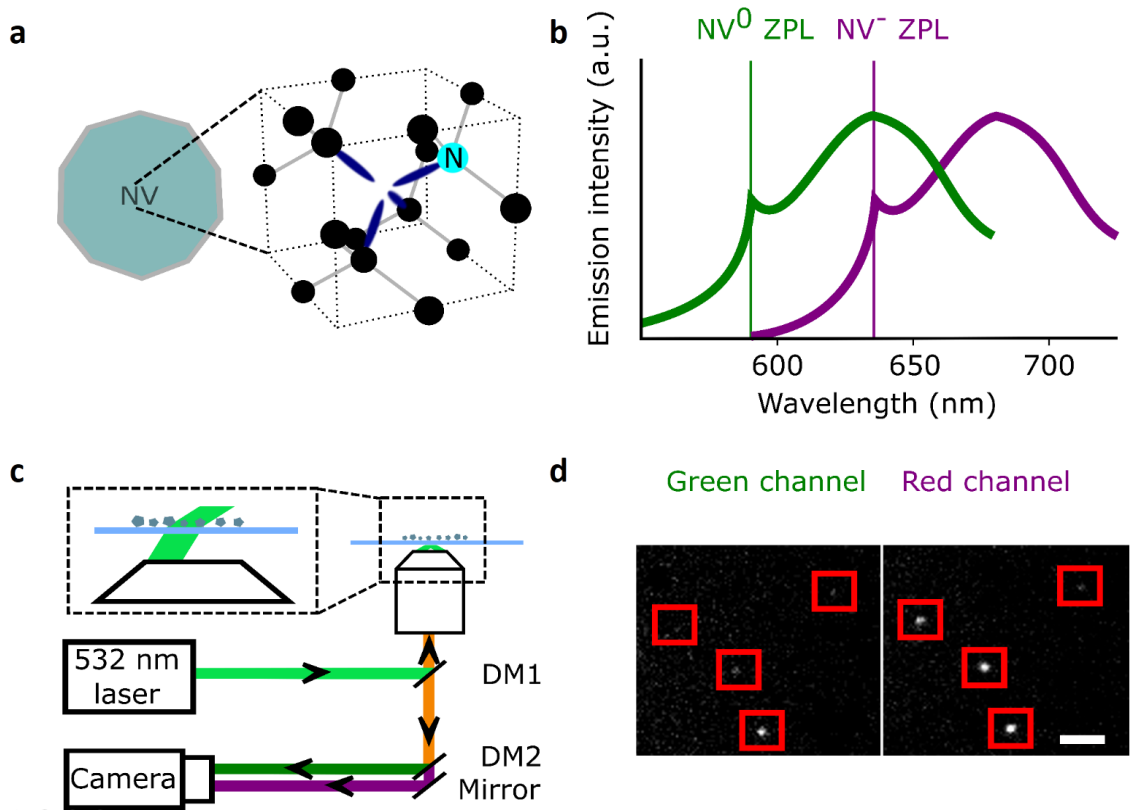

**Figure 1:** wide-field imaging of NDs. a, Schematic representation of a ND containing a NV center including the atomic structure of the defect. b, Emission spectrum of NV$^-$ measured in the 44-nm NDs and a schematic of the blue shifted emission of the neutral state (ZPL: zero-phonon line). c, Simplified representation of the wide-field microscope with NDs immobilised on the microscope slide. The zoomed-in region shows variable-angle illumination; DM1: dichroic with reflective bands at 532 and 638 nm; DM2: 650 nm long pass dichroic. d, Section of the field of view with fluorescent NDs. 532 nm excitation at 7.8 kW/cm$^2$, 100 ms exposure; the red squares represent the NDs localised by our software (Gapviewer), the photon count for each ND per camera exposure is calculated from a region of interest covering the fluorescent spot; scale bar = 8 μm.

**Results**

**High-throughput analysis of different ND samples.** To study the heterogeneity of ND photophysics, we characterised samples of different sizes (5-200 nm), different manufacturing processes and different NV center content (see *Methods*). The 5 and 10-nm diameter NDs are usually too small to contain more than 2 NVs per ND and have only up to ~1% of NDs with NV center.[28] Since they are undoped, the 50-nm diameter NDs are expected to have up to ~10 % of fluorescent NDs containing only 1-2 NV centers per particle.[21] The doped 40 and 44-nm diameter NDs contain up to 4 NV/ND with a larger fraction of bright NDs (up to 70 % for the 40-nm, see *Methods*). Finally, the 200-nm doped NDs are all expected to be emitting fluorescence and can have up to 100 NV/ND because of their larger size.

We studied the ND samples using a wide-field two-channel single-molecule fluorescence microscope (see *Methods*). Our large field of view (50 x 80 μm) allowed us to use a low ND density (down to one fluorescent ND for 40 μm$^2$, Figure 1d), ensuring that we are observing single NDs (Figure S1). We collected 96 to 589 time traces per sample and used all data points to build the photon count distribution of the sum of red and green channels. To detect changes in the emission spectrum, we compared the relative intensities of the red and green channels using the R/G ratio (*R/G ratio=Spot Intensity$_{red}$/(Spot Intensity$_{red}$ + Spot Intensity$_{green}$)*). Since the emission of NV$^0$ is blue shifted (~60 nm) compared to NV$^-$, this ratio offers us a facile way to study the charge of the NV center.

Our studies of the 50-nm undoped NDs showed that most time traces showed a total photon count of 1000/100 ms and a R/G ratio of 0.9 (Figure 2a, left; Figure 2b). Other traces are brighter (1500 photons/100 ms) with a lower R/G ratio (0.6), as more light is detected in the green channel (Figure 2a, right). These two populations with distinct R/G ratio can also be seen as a very broad R/G ratio distribution on a 2D histogram (Figure 2b). We attributed these two R/G ratio states to the different charge states of the NV center (0.6 for $NV^0$ and 0.9 for $NV^-$) based on their respective emission spectra and our detection efficiency for the red and green channels (Figure S2). The photon counts we obtained are 20-fold higher than previously reported for a single NV center using wide-field imaging.[29]

The 2D histogram for 44-nm NDs shows the same main population as the 50-nm undoped NDs (Figure 2c); this similarity allows us to assign this intensity to a single $NV^-$ emission, since this sample was manufactured to contain 1NV/ND; we further verified the presence of single $NV^-$ per ND in the 44nm NDs using photon correlation experiments (Figure S3). Notably, no clear $NV^0$ signal at a R/G ratio of 0.6 is observed in these 44-nm NDs, which confirms that the population with a 0.6 R/G ratio in Figure 2b is $NV^0$. The proportion of $NV^-$ in doped NDs is expected to be high (*e.g.* >65 % in 40-nm NDs according to the supplier) as ND doping involves high concentration of nitrogen (up to 200 ppm) to improve the probability of NV center formation; high nitrogen content is known to stabilize NV's negative charge state as nitrogen acts as an electron donor for NV centers.[26] Similar distributions of photon count and R/G ratio were measured in 40-nm doped NDs (Figure 3a-b, row 5), an expected result based on the comparable size and manufacturing process for the 40 and 44-nm doped NDs.

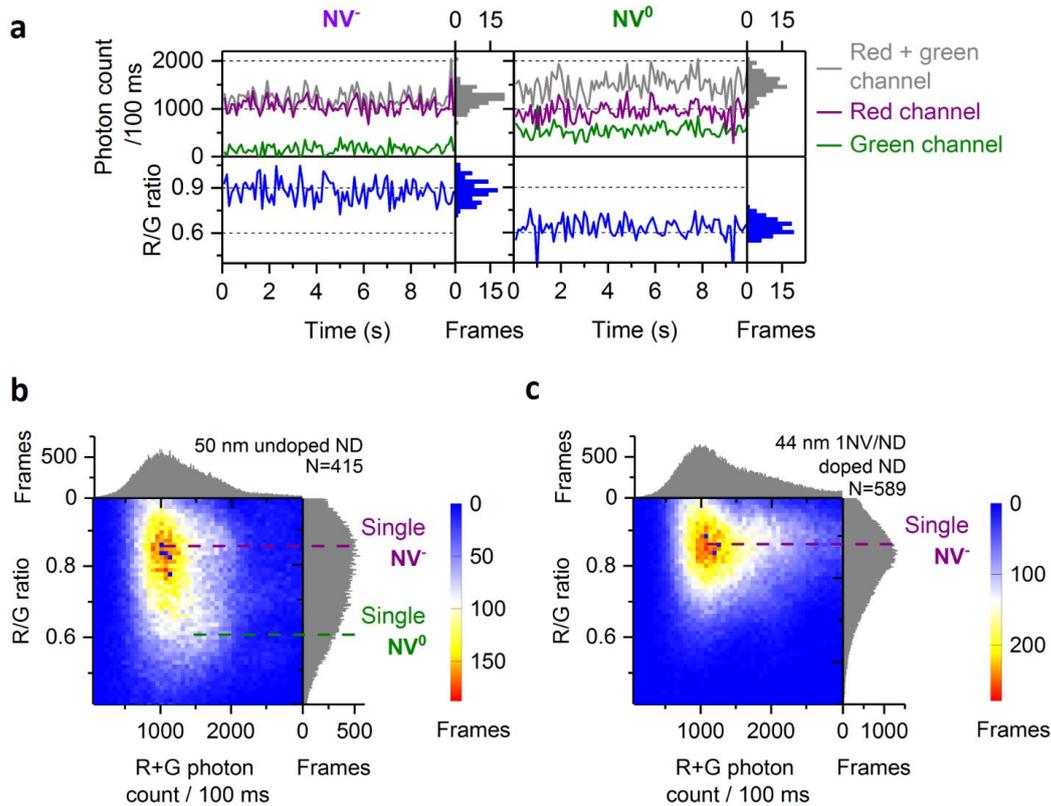

**Figure 2:** charge states of NV center in doped and undoped NDs. a, Typical times traces from 50-nm undoped NDs corresponding to single NV⁻ and NV⁰. b, 2D histogram of photon count and R/G ratio in 50-nm undoped NDs; purple and green dotted lines indicate respectively single NV⁻ or NV⁰ fluorescence. c, 2D histogram of photon count and R/G ratio in 44-nm 1NV/ND doped NDs; the purple line indicates the fluorescence from single NV⁻. For figure b and c, N is the number of fluorescent NDs observed. For all the figures, "Frames" is the frequency of the measurement from a ND (see Figure 1 for details). The immobilized NDs were imaged in air using 532 nm excitation 7.8 kW/cm$^2$ and 100 ms exposure for 25 s.

We then examined sub-20-nm NDs, which are challenging samples as they contain more unstable NV charge states.[28, 30] In such small particles, the fluorescent defect is closer to charge traps at the surface that can act as electron acceptors to NV$^-$.[31] We observed that the photon count of the 10-nm doped NDs mainly originates from single NV centers, which is expected given their small size (Figure 3a row 3); the distribution of the R/G ratio is centered around 0.7 (Figure 3b row 3), suggesting that the charge conversion may happen within 100 ms. We attribute this charge state instability to the NV center's proximity to the surface.

The 5-nm NDs show low brightness, a small bright fraction (<1 %; see Methods) and, unlike all the other samples, most of the 5-nm NDs photobleach within a few seconds. Nonetheless, we can confirm the rare presence of stable and bright NDs (>2000 photons/100 ms; Figure 3a row 2), as previously reported.[28, 30] Given the scarcity of such emitters (<0.1% of the total 5-nm NDs) one cannot exclude the possibility that we detected fluorescence from a small subpopulation of NDs having a larger size than 5 nm (*e.g.*, 10-30 nm, as reported by Vlasov *et al.*).[30]

The 200-nm doped NDs exhibit a very broad photon count distribution (Figure 3, bottom; showing only NDs with photon count <5000 photons/100 ms (~20% of the 200-nm NDs)). The first maximum of the photon count distribution is above 2000 photons/100 ms, indicating that the sample contains no single NVs per ND, consistent with the probability of having NDs containing one NV center being very low in such large doped NDs. Finally, no NV$^0$ fluorescence is present in the R/G distribution (Figure 3b bottom), very likely due to their size and high nitrogen concentration.

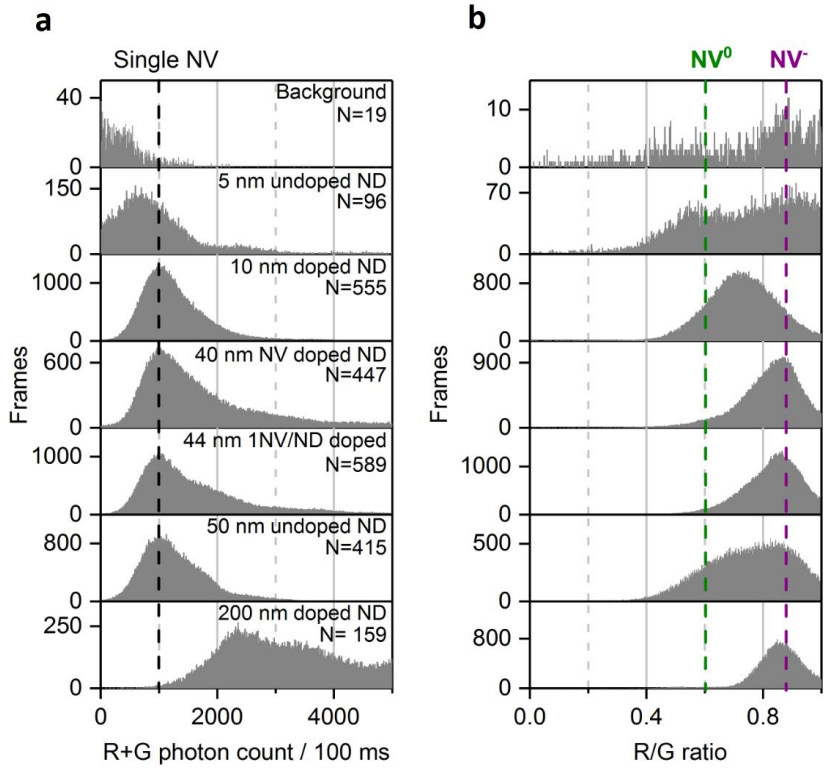

**Figure 3:** distribution of photon count (a) and R/G ratio (b) in different NDs samples. Black dotted line marks the photon count collected from single NV center. The background signal is the fluorescence originating from impurities in the microscope slide. The purple and green dotted lines indicate respectively $NV^-$ and $NV^0$ fluorescence. Using multiple-Gaussian fitting, the proportion of single NV center ($NV^-$ or $NV^0$) per ND is estimated be > 70 % for the 50-nm undoped and 10-nm doped NDs, < 50 % for the 44 and 40 nm-doped NDs and 0 % for the 200-nm doped NDs. The same analysis for the 5-nm undoped NDs is not possible because of their low brightness and low statistics. N is the number of fluorescent NDs observed and "Frames" is the frequency of the measurement from a ND (see Figure 1 for details). The immobilized NDs were imaged in air with 532 nm excitation 7.8 kW/cm$^2$ 100 ms exposure for 25 s.

**Dynamic behavior of NDs.** To capture dynamic transitions in NDs occurring in the timescale of seconds, we studied 40-nm NDs exhibiting NV charge-state transitions. For the population with single NVs, we measured steady fluorescence in 60% of them (Figure 4a) as opposed to the remaining 40%, in which clear dynamic behavior was observed. Dynamic traces are mostly due to NV centers with unstable charge states (Figure 4b (i) and (ii)), which causes in rare cases some on/off blinking (<1 % of the total traces, Figure 4b (iii)). According to previous studies, fluctuations in ND fluorescence is due to the proximity (<10 nm) between the NV center and the ND's surface;[24, 32] electron transfer from the NV center to surface charge traps allows the NV defect to switch from $NV^-$ to $NV^0$ and then to the non-photoactive $NV^+$ charge state, causing blinking.

To study the dynamic traces, we first defined the charge states in terms of photon count and R/G ratio using stable time traces (Figure S4). We assigned emitters having a photon count of 4000/s and a R/G ratio of 0.6 to single $NV^0$ (16% of static traces, Figure 4a top). We also assigned the traces showing a photon count between 2000 and 4000 photons/s and a R/G ratio of 0.7 to 0.8 to time-averaged values of the $NV^-$ and $NV^0$ states (38% of static traces, Figure 4a middle); such averaging is consistent with charge conversion in bulk diamonds that may occur within the μs timescale.[33] Finally, we assigned the NDs emitting 2500 photons/s with a R/G ratio of 0.9 to single $NV^-$ (48 % of static traces; Figure 4a bottom).

We then collected dynamic and long traces (lasting 30-60 min), which provided enough statistics to compare state transitions within NDs ($N = 32$, providing >600 dwells). The dynamic traces showed either two-state or three-state transitions (Figure 4b (i) and (ii)) with the transition frequency varying among NDs (Figure S5). To investigate if these dynamic NDs share similar states and dwell times, we performed Hidden Markov modelling (HMM) analysis, which showed that a three-state model was sufficient to fit our dynamic traces (Figure S8);[34] the three states identified ($NV^-$ at 0.85, $NV^0$ at 0.67 and a time-averaged state at 0.75, Figure 4c) correspond well to those seen in static traces. The slight difference (+/-0.07) between the values from the static and dynamic traces is likely due to small errors in the state allocation by HMM.

We used the dwell times from the HMM analysis to calculate the lifetimes of each NV charge state. Dwell time distributions from the $NV^-$ and $NV^-/NV^0$ states were fitted with a single-exponential decay function (Figure 4b middle and bottom histograms) while the dwell time distribution from the $NV^0$ state had to be fitted with a double exponential decay function (Figure 4b top histogram) as the single exponential decay was clearly missing a long-dwell component. Most of the $NV^0$ dwells are less than 10 s as showed by the amplitude (A) of the lifetimes ($\tau$): $\tau_1$=3 s $A_{\tau1}$= 90 %; $\tau_2$=53 s $A_{\tau2}$=10%. The most representative $NV^0$ lifetime is shorter than the two other states ($\tau$= 57 s for $NV^-/NV^0$ and $\tau$=38 s for $NV^-$, see Figure 4d), which is expected when using 532 nm excitation, a wavelength known to pump $NV^0$ back to $NV^-$, thus making $NV^0$ lifetime shorter in our imaging conditions.[35, 36]

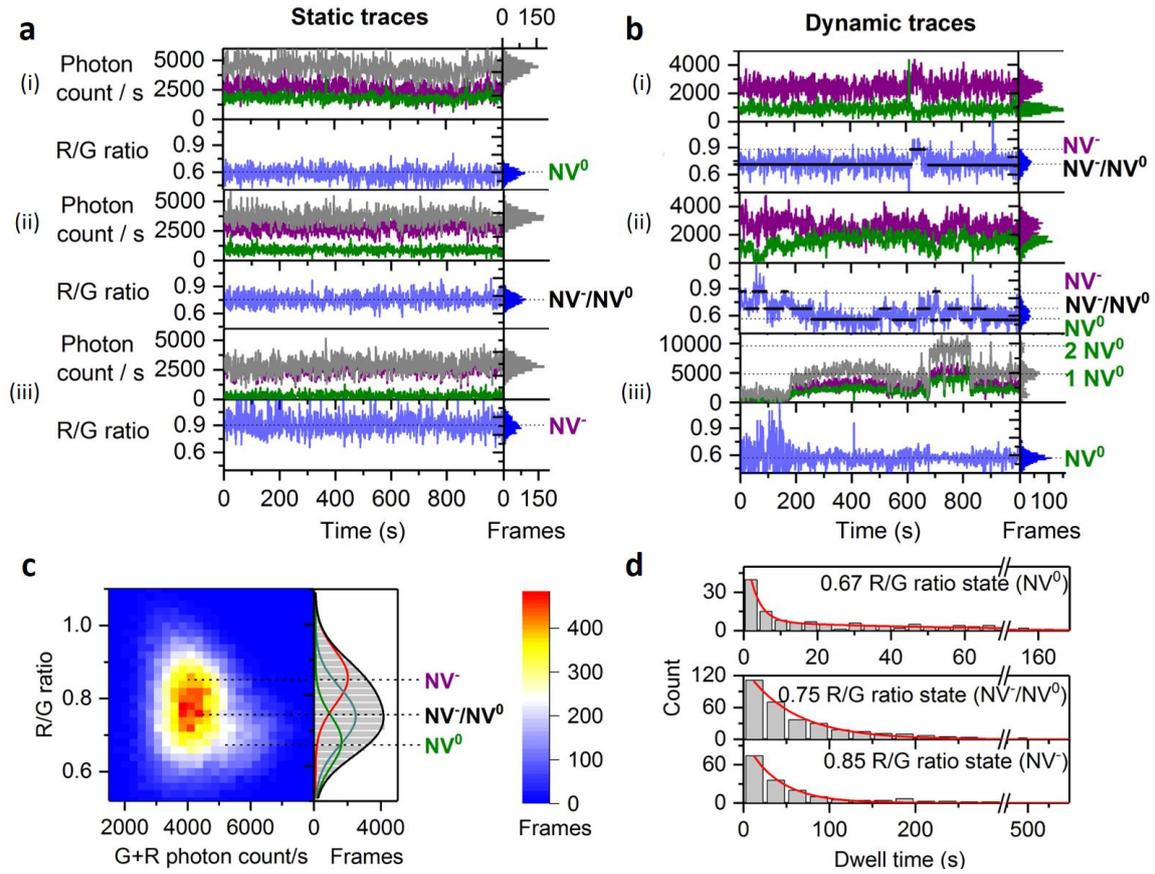

**Figure 4:** states and dynamics of NV center in NDs. a, Typical static time traces corresponding to the two different charge states of single NV centers ((i): $NV^0$ R/G ratio = 0.6 – 4000 R+G photons/s; (iii): $NV^-$: R/G ratio = 0.9-2000 R+G photons/s) and a time-average of the two charge states ((ii): $NV^-/NV^0$: R/G ratio = 0.7-0.8 - 2000-4000 R+G photons/s). Red and green photon counts are shown in purple and green, respectively; the sum of red and green photon counts is in gray. b, Typical dynamic traces; the first trace (i) shows two-state transitions as demonstrated by the spectral inversion; the second trace (ii) is more dynamic and shows three-state transitions; the last trace (iii) shows two-level blinking c, 2D histogram of 32 dynamic traces with their R/G ratio distribution on the right side of the figure; the R/G distribution is fitted with 3 Gaussian profiles centered at the R/G ratio values calculated by HMM modelling, which fits well the distribution and the 2D histogram especially the brighter $NV^0$ population. Gaussian centers: $NV^0$ R/G ratio = 0.67; time-averaged $NV^-/NV^0$: R/G ratio

= 0.75; NV$^-$: R/G ratio = 0.85. d, Dwell time analysis of the 3 states by HMM. The distribution of dwell times is fitted with a double exponential decay function for the NV$^0$ state (top histogram, fitting function shown in red). The lifetimes are $\tau_1$=3 s +/- 0.3 s $A_{\tau 1}$= 90 %; $\tau_2$=53 s +/- 7.7 s $A_{\tau 2}$=10% for NV$^0$. The distribution of dwell times is fitted with a single exponential decay function for the NV$^-$/NV$^0$ and NV$^-$ states (middle and bottom histograms). The lifetimes are: $\tau$= 56 s +/- 1.7 s for NV$^-$/NV$^0$; $\tau$= 38 s +/- 1.5 s for NV$^-$. "Frames" is the frequency of the measurement from a ND (see Figure 1 for details). The immobilized NDs were imaged in air using 532 nm excitation 3.4 kW/cm$^2$ and 1 s exposure for 20 min to 60 min.

**Charge state manipulation using pH.** Since we could detect different charge states of the NV center, we investigated our ability to modify the ND's emission by immersing immobilized 10-nm NDs into separate solutions containing DNA, proteins, and a reducing agent (DTT); none of these solutions showed detectable effects on the photon count or the R/G ratio (Figure S10). Nevertheless, changes in pH were found to have a clear effect on the NV center's charge (Figure 5a and b), where high pH shifts the R/G distribution towards higher values (from 0.7 to 0.8 R/G ratio in Figure 5a).

To explore the possibility that these NDs could be used as pH sensors, the reversibility of the effect was tested, by imaging the same field of view (containing ~140 NDs) after consecutive immersions into acidic and basic solutions. The modes of the R/G ratio distribution (Figure 5c) clearly showed that the effect of pH on the R/G ratio is indeed reversible.

We have also examined whether we have the resolution to detect the increase of the R/G in basic pH on the same ND particle; this was indeed the case (see example in Figure 5d). In addition to this change in R/G ratio, there was also a decrease of photon count (from 1750 to 1250 photons/100 ms) at pH 12.8. Such increase of R/G ratio and decrease of brightness correlates well with our observation of charge state conversion on the 50-nm undoped and 40-nm doped NDs (Figure 2; Figure 4), with the deviations of the R/G ratio values from 0.6 and 0.9 being attributed to the charge instability in the 10-nm NDs. However, our data indicates that the NV center will stay longer in a charge state ($NV^0$ for acidic pH or $NV^-$ for basic pH) within 100 ms, leading to changes in the R/G ratio.

The mechanism behind this pH-dependent charge transition of NV probably is likely to involve deprotonation of ND surface groups (mainly OH or COOH groups generated during ND acid cleaning, and affecting the ND's surface charge), thus creating more negative charges around the NV center.[37] Our findings confirm previous reports that NDs photophysics is affected by surface chemistry and demonstrate that they can be used for pH nanosensing.[13,27]

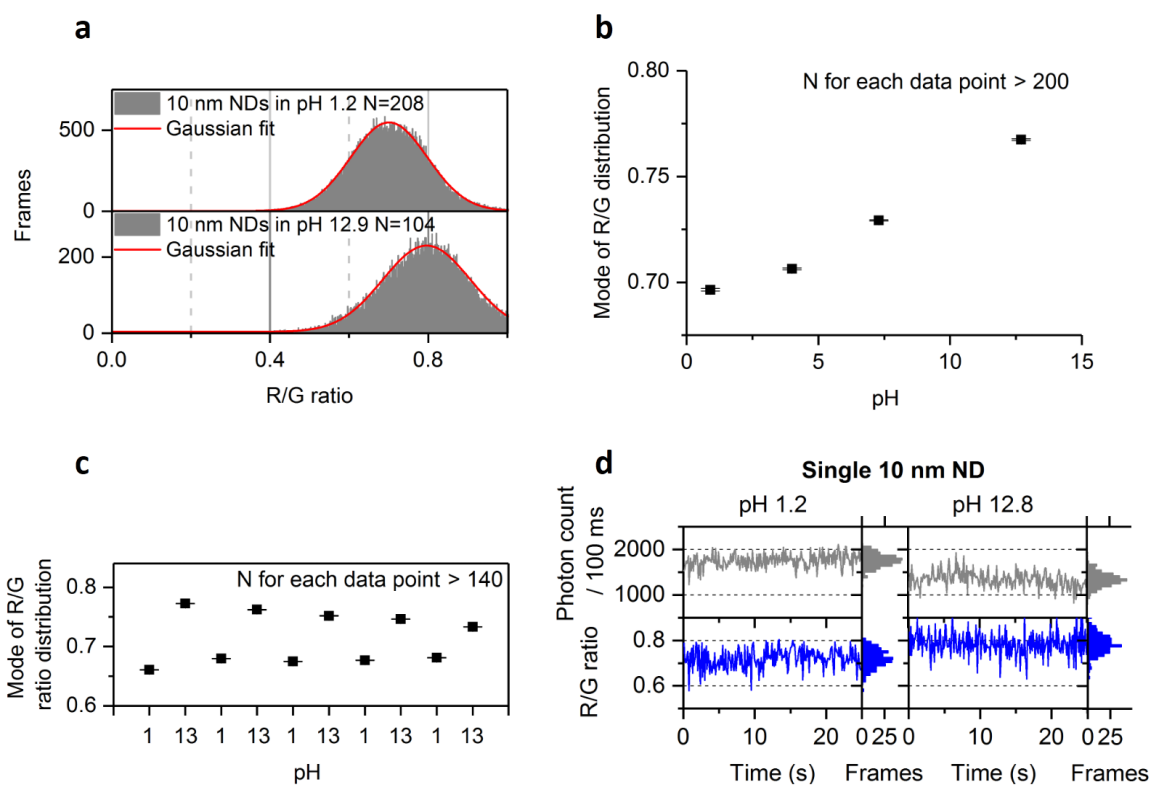

**Figure 5:** effect of pH on 10-nm NDs. a, Increase of the R/G ratio distribution in the 10-nm doped NDs at high pH. Aqueous solutions used: 0.01 M HCl for pH 1.2, and 0.1 M NaOH for pH 12.8. b, Relation of the mode of the R/G ratio distribution with pH following Gaussian fitting of the distribution (0.01 M HCl for pH 1.2, 0.1 M NaOH for pH 12.8 and commercial buffer solutions used for the other pH data points). c, Reversibility of the effect following multiple washing on the same field of view containing more than 140 NDs (washing was performed using deionized and filtered water; the same acidic/basic solutions as in figure a were used). The convergence of the modes after the first repeat may be due to the washing steps that gradually become less efficient. d, Observation of the charge state conversion on the same ND particle in acidic and basic solutions (same solutions used as in Figure a). R+G photon count is shown here. Approximately 75% of the single NV centers observed showed

this increase of the R/G ratio in basic solution. The error bars (Figure 5b and c) indicate the standard error from the Gaussian fitting. N is the number of fluorescent NDs observed and "Frames" is the frequency of the measurement from a ND (see Figure 1 for details). The immobilized NDs were imaged with 532 nm excitation 7.8 kW/cm$^2$ 100 ms exposure for 25 s.

**Discussion**

Our work demonstrated that wide-field imaging and automated time-trace analysis is a powerful approach to characterize ensembles of NDs. By examining large numbers of single NDs in parallel, we were able to detect single NVs in up to 500 NDs per sample, providing helpful comparisons of the proportion of single emitters in different samples based on their photon count distribution, a task facilitated by the random orientations of the spin-coated NDs on the microscope slide. We also introduce the use of the R/G ratio to analyze the charge state of the NV center, a crucial determinant on the ND photophysical behavior. Our technique can be easily implemented to screen different NDs samples to investigate the proportion of single NV/ND and the charge stability of the NV center since measuring >100 NDs takes only seconds.

Since we used a statistical approach to define the photon count provided by a single NV, it is difficult to confirm that a given ND contains only one NV center based on a single observation if no charge transitions are observed. For this reason, photon correlation experiments are more suited in this case.[23] However, this limitation could be overcome by investigating the orientation of NV center using defocused orientation and position imaging.[38]

Our results on the charge state instability in 10-nm ND or undoped 50-nm ND confirm previous reports that established the negative impact of ND size or low nitrogen concentration on the stability of the $NV^-$.[24-26] Further, our high-throughput approach allowed studies of small but significant subpopulations of the NDs, such as the one having dynamic single NV centers; based on 32 dynamic time traces, we could estimate the lifetimes of the $NV^-$ and $NV^0$ in NDs. Such a study was not reported before as confocal or wide-field measurements described lifetimes of $NV^-$ from only a single particle.[29, 31]

The lifetimes we measured for $NV^-$ and $NV^0$ (from 3-60 s) significantly broadens our understanding of dynamic NDs, as previous studies only report on the on and off lifetimes of NDs. Indeed, measuring blinking might be easier to perform in NDs, but since the NV center has two fluorescent states (NV$^-$ and NV$^0$) and one non-fluorescent state (NV$^+$), blinking-based studies are ambiguous with regards to the charge transitions involved in the charge-transfer events. The closest report to our results is a study published by Aslam *et al.* who investigated charge state transitions of a single NV center in bulk diamond, and measured lifetimes for NV$^-$ and NV$^0$ (57 ms and 465 ms respectively) three orders of magnitude shorter than what we measured; however, those measurements used significantly different conditions (study in bulk diamond instead of in a ND, and use of a different excitation wavelength (593 nm), and examined only one NV center).[35] The presence of a longer component only for NV$^0$ might be due to sub-seconds transitions into the dark NV$^+$ state that could slow down the transition from NV$^0$ to NV$^-$.

We also showed that NDs can be directly used for pH nanosensing. In ways similar to the studies of Petráková *et al.* and Karaveli *et al.*, who showed that the NV charge state in NDs can be used for sensing (electrical potential or surface chemistry changes), we demonstrated a simple practical implementation of ND sensing without the need for functionalization or equipment such a spectrograph or microwave generator for spin state manipulation.[13, 14, 27] Notably, pH did not significantly impact the R/G ratio in the 44-nm doped NDs (Figure S9), consistent with the prediction by Petráková *et al.* in 2012, which proposed that only small particles (10-20 nm) could lead to optically detectable transitions of NV's charge because of required proximity to the surface.[13]

**Conclusion**

In summary, our ability to detect simultaneously hundreds of single NV centers and their charge states is a powerful screening method for material scientists who manufacture NDs and need a facile and reliable way to characterize their photophysical properties. Such analysis will also interest physicists working on quantum optics applications, who need to screen ND samples for brightness, charge stability, and the fraction of NDs having a single NV center. Our capability to study dynamic NDs directly and in parallel will enhance our fundamental understanding of NV charge transitions in a nanocrystal and our ability to maintain the negative charge state, which is the charge state that is mostly used in ND commercial applications. Considering that NDs show great photostability compared to organic fluorophores, our findings should foster more applications of single-molecule fluorescence imaging and tracking experiments *in vitro* and in living cells, as well as sensing experiments such as pH monitoring in microfluidics or pH mapping inside biological samples.[39] Finally, our method should also facilitate the development of biosensing assays based on the measurement of the NV charge state conversion and its dynamic behavior. Such assays will be helped by further study on the effect of bio-functionalization on the charge state since the NV center can be affected by the surface chemistry.

**Materials and Methods**

*Nanodiamond samples:* Doped nanodiamonds of 10, 40, 44 and 200-nm in diameter were commercial high-pressure-high temperature (HPHT) samples enriched in nitrogen followed by particle irradiation to generate vacancies (providers: 10 and 40-nm: Adámas nano; 44-nm: FND biotech; 200-nm: Columbus Nanoworks). Following the high-temperature annealing of NDs, the vacancies recombine with the nitrogen to form NV centers. When not provided by the manufacturer, the fraction of NDs containing NV was coarsely estimated by dividing the density of fluorescent NDs by the density of NDs deposited on the glass surface. The values obtained from 2 to <0.1 % are within the range of values reported in the literature (0.03 to 70%) depending on the size and manufacturing process.[4, 28] The 44-nm NDs were manufactured to contain a maximum proportion of one NV center per ND. The NV center's emission spectra was measured in all our samples (see example in Figure S11). The presence of single quantum emitters in the 44-nm sample was confirmed by photon correlation experiments (Figure S3). 50-nm NDs were undoped HPHT particles. All the samples were acid cleaned, sonicated and their size distribution was confirmed by single-particle tracking and/or dynamic light scattering. The 10-nm doped ND have a nominal size of 10 nm but our analysis revealed a size distribution centered at 20 nm; this discrepancy may be due to the irregular shapes of small NDs (*i.e.* flakes). [40]

*Microscope and imaging:* The particles were spin coated at a very low density (down to 1 fluorescent ND for 40 μm$^2$) and imaged using a single-molecule desktop wide-field microscope Nanoimager S (*Oxford Nanoimaging*) with a 1.4 NA oil immersion objective. The emitted light is split into a green and red imaging channels with a long pass filter at 650 nm for the red channel (see Figure 1c). The 1W 532 nm CW laser allows us to detect single NV centers with 10-1000 ms time resolution when used at full intensity (7.8 kW/cm$^2$). The exposure time of 100 ms was selected for most experiments as it provided the best SNR for a minimal acquisition time (25 s). Illumination for the TIRF objective was at ~50° to remove out-of-focus background. All the samples were exposed to maximum excitation intensity for 30-60 sec to photobleach other emitting species (*e.g.*, surface defects). The photon count distribution from 80 to 300 particles per sample was collected by imaging different fields of view. A custom-built confocal set-up was used to perform spectral measurement and photon correlation experiments as we previously reported.[41]

For pH sensing, the 10-nm NDs were immersed in different buffer solutions using silicon gasket to form wells. A 0.01 M HCl solution was used for pH 1.2, a 0.1 M NaOH solution for pH 12.8, a 1X phosphate-buffered saline solution (PBS) for pH 7 and the pH 4 measurement was carried out using a commercial buffer solution (Hanna instruments). Washing was performed using deionized and filtered water (220 nm pores) and the acquisition was done 1-3 min following the solution addition.

*Time trace analysis and HMM:* The raw image was processed by home-built software (GapViewer) that detects diffraction-limited spots by their intensity. It performs background subtraction for each frame using the intensity around the Gaussian profile of the NDs emission. The photon count is calculated by adding both channels and the R/G ratio was computed as $\frac{Spot\ Intensity_{red}}{Spot\ Intensity_{red} + Spot\ Intensity_{green}}$. Based on the NV center emission spectra of the two charge states previously reported[35] and the wavelength dependence of the quantum efficiency from the microscope's sCMOS camera (Figure S2), we expect a R/G ratio difference of 0.3 between $NV^-$ and $NV^0$ with $NV^0$ being brighter than $NV^-$. The dynamic traces were manually selected based on their R/G ratio (within 0.6 and 0.9) and their intensity (0-1500 photons/s for the green channel and 1000-3200 photons/s for the red channel). Some traces (<10%) were excluded from the analysis if their R/G ratio transitions were showing more than 3 states because it could indicate the presence of a second but dimmer NV center. Following HMM processing (EbFRET, 10 restarts, 0.001 precision and prior strength set with a 0.2 center (priors set only to extract dwell times)), the durations of the dwell times for each state was extracted and fitted with a double exponential decay.

AUTHOR INFORMATION

**Corresponding Author**

Prof. Achillefs N. Kapanidis

**Notes**

A.N.K is a shareholder and consultant in Oxford Nanoimaging, a company that designs, manufactures and supports single-molecule fluorescence microscopes.

ACKNOWLEDGMENT

We thank Ella Bentin and Dr. Abhishek Mazumder for helpful discussions and the Micron Oxford Advanced Bioimaging facility. This work was supported by the EPSRC Centre for Doctoral Training in Diamond Science and Technology (EP/L015315/1) (to M.S.) as well as the Wellcome Trust grant 110164/Z/15/Z and UK BBSRC grant BB/N018656/1 (to A.N.K.). O.A.W gratefully acknowledges support by the European Research Council under the EU Consolidator Grant 'SUPERNEMS'.

ABBREVIATIONS

NV, nitrogen-vacancy; ND, nanodiamonds; ZPL, zero-phonon line; DM, dichroic mirror; R/G, red/green; R+G, red + green; HMM, Hidden Markov modelling; A, amplitude; DTT, dithiothreitol; HPHT, high-pressure-high temperature; PBS, phosphate-buffered saline solution; sCMOS, scientific complementary metal-oxide-semiconductor.

**Supporting information**

**High-Throughput Detection and Manipulation of Single Nitrogen-Vacancy Center's Charge in Nanodiamonds**

*Maabur Sow, Horst Steuer, Sanmi Adekanye, Laia Ginés, Soumen Mandal, Barak Gilboa, Oliver A. Williams, Jason M. Smith, Achillefs N. Kapanidis\**

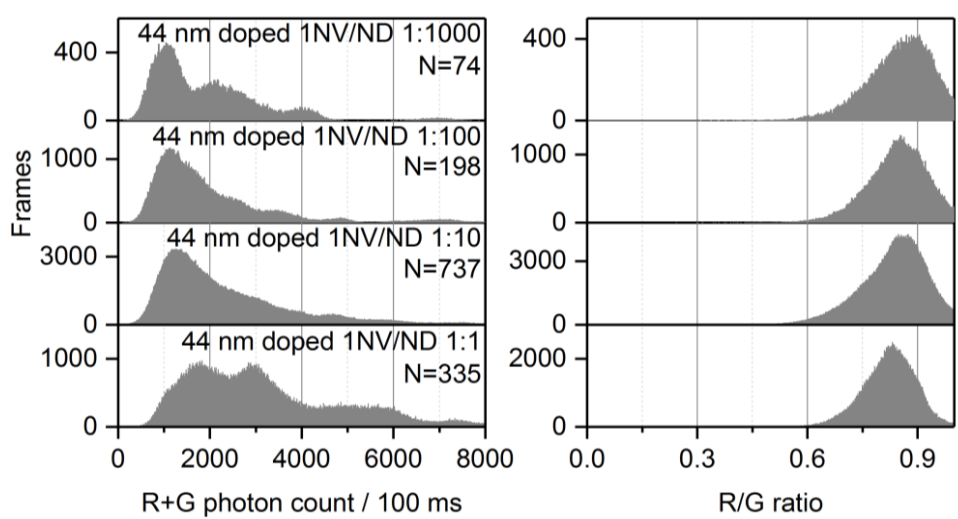

**Figure S1: distribution of photon count and R/G ratio in different NDs dilutions.** We previously checked that the NDs solutions are not aggregated by dynamic-light scattering or single-particle tracking analysis. However, there is a possibility that aggregates may have formed when NDs were deposited on the slide, which could make the first mode of the photon count distribution (*i.e.* 1000 photons/100 ms from Figure 3a) originating from two or more NV centers instead of a single NV. To test this hypothesis, we measured the distribution of the photon count from different dilutions of the 44-nm 1NV/ND ND solution. If the mode at 1000 photons/100 ms is due to aggregates, the highest dilution (1:1000) should show a mode inferior to 1000 photons/100 ms. Results in the figure above shows that the most diluted sample (1:1000, see top row) still has a clear peak at 1000 photon/100 ms while the undiluted

sample (1:1) displays a first maxima at 2000 photon/100 ms (see bottom row). The fact that this mode at 1000 photons/100 ms becomes even more distinct at a low dilution is consistent with the data from the 10 and 50-nm NDs, which support our previous deduction that the photon count of 1000 photons/100 ms corresponds to the emission of single NV$^-$ center. N is the number of fluorescent NDs observed and "Frames" is the frequency of the measurement from a ND (see Figure 1 for details). The 44-nm doped NDs (1NV/ND) were diluted in deionized water and imaged in air with 532 nm excitation 7.8 kW cm$^{-2}$ with 100 ms exposure for 25 s.

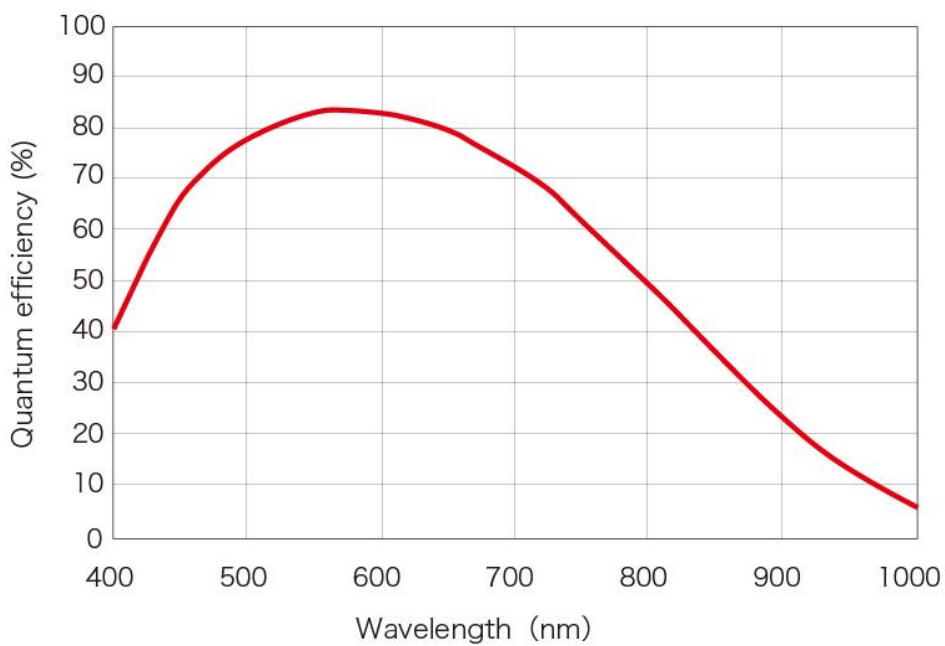

**Figure S2: quantum efficiency of the scientific CMOS camera implemented in the wide-field microscope.** The quantum efficiency of the sCMOS is at its maximum (>80%) between 520 and 650 nm which corresponds to the main emission band of $NV^0$. Such a difference in quantum efficiency could explain why we measured a higher total photon count (green + red channel) of $NV^0$ 50 % superior to $NV^-$. Data obtained from the website of the sCMOS' manufacturer on 1-jan-19 (https://www.hamamatsu.com/jp/en/C13440-20CU.html#1328482115119).

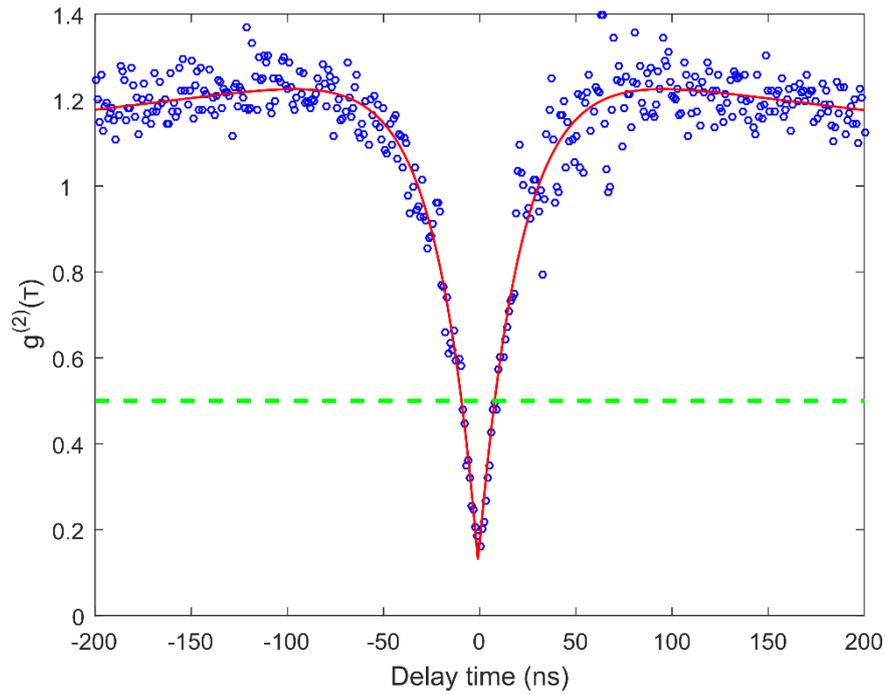

**Figure S3: second order correlation function from single NV center in a 44-nm doped NDs (1NV/ND).** The presence of single NV center was also confirmed by photon correlation experiments. The drop of the second order correlation function (*i.e.* antibunching) below 0.5 confirms that the emitter observed is a single photon source. Antibunching confirming the presence of single NV center was observed in 23 % of the NDs measured (N=311, see example above).

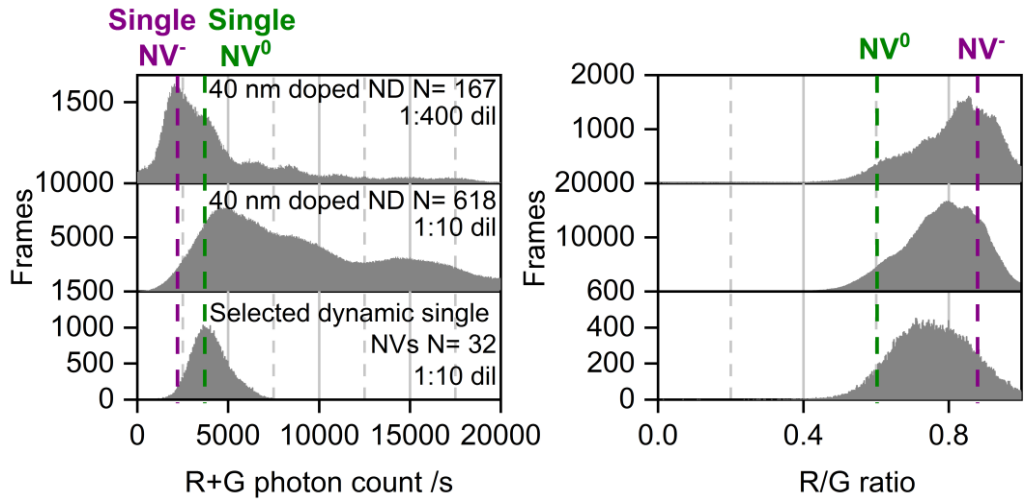

**Figure S4: distribution of photon count and R/G ratio in different NDs samples.** For the study of dynamic NDs, the nanoparticles were imaged with 1 s exposure time and a reduced laser excitation (3.4 kW cm$^{-2}$) for 20 to 60 min. The photon count distribution of the diluted 40-nm doped NDs (1:400 see top row) allowed us to identify the level from single NV$^-$ and NV$^0$ as described previously in the first part of the result section (see Figure 2 and 3). To increase the probability to obtain time traces of dynamic single NV centers per field of view, we increased the ND concentration (see middle row, dilution 1:10). Time traces corresponding to single NV centers (dynamic and stable ones) were manually selected based as described in materials and methods. The selected dynamic single NVs (see bottom row) contains mainly time-averaged NV$^-$/NV$^0$ and NV$^0$ as shown by the R/G ratio distributions. The reason why most of stable NV$^-$ traces were excluded is that the intensity of NV$^-$ in the green channel was approaching zero, which was causing error in the HMM analysis. Nevertheless, the dynamic ones provided enough dwells in the NV$^-$ state for the HMM algorithm to detect and populate this state as shown in Figure 4c. N is the number of fluorescent NDs observed and "Frames" is the frequency of the measurement from a ND (see Figure 1 for details).

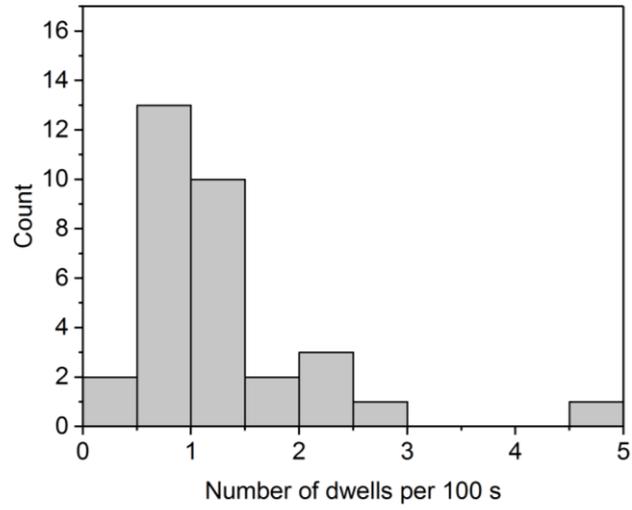

**Figure S5: distribution of the number of dwells 100/s from the dynamic time traces.**

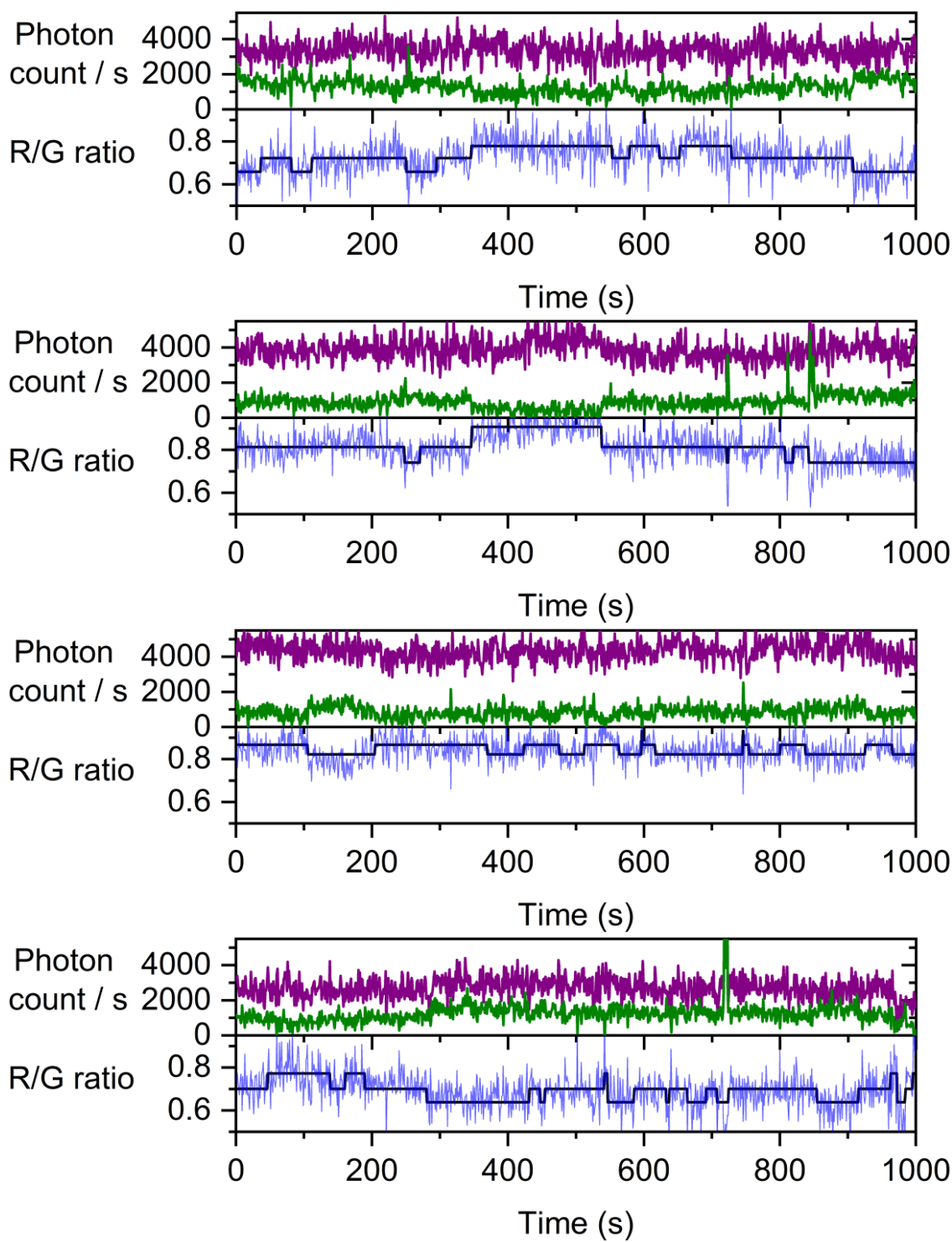

**Figure S6: Typical moderate-fluctuation dynamic traces.** The immobilized NDs were imaged in air using 532 nm excitation 3.4 kW cm$^{-2}$ and 1 s exposure for 20 min to 60 min.

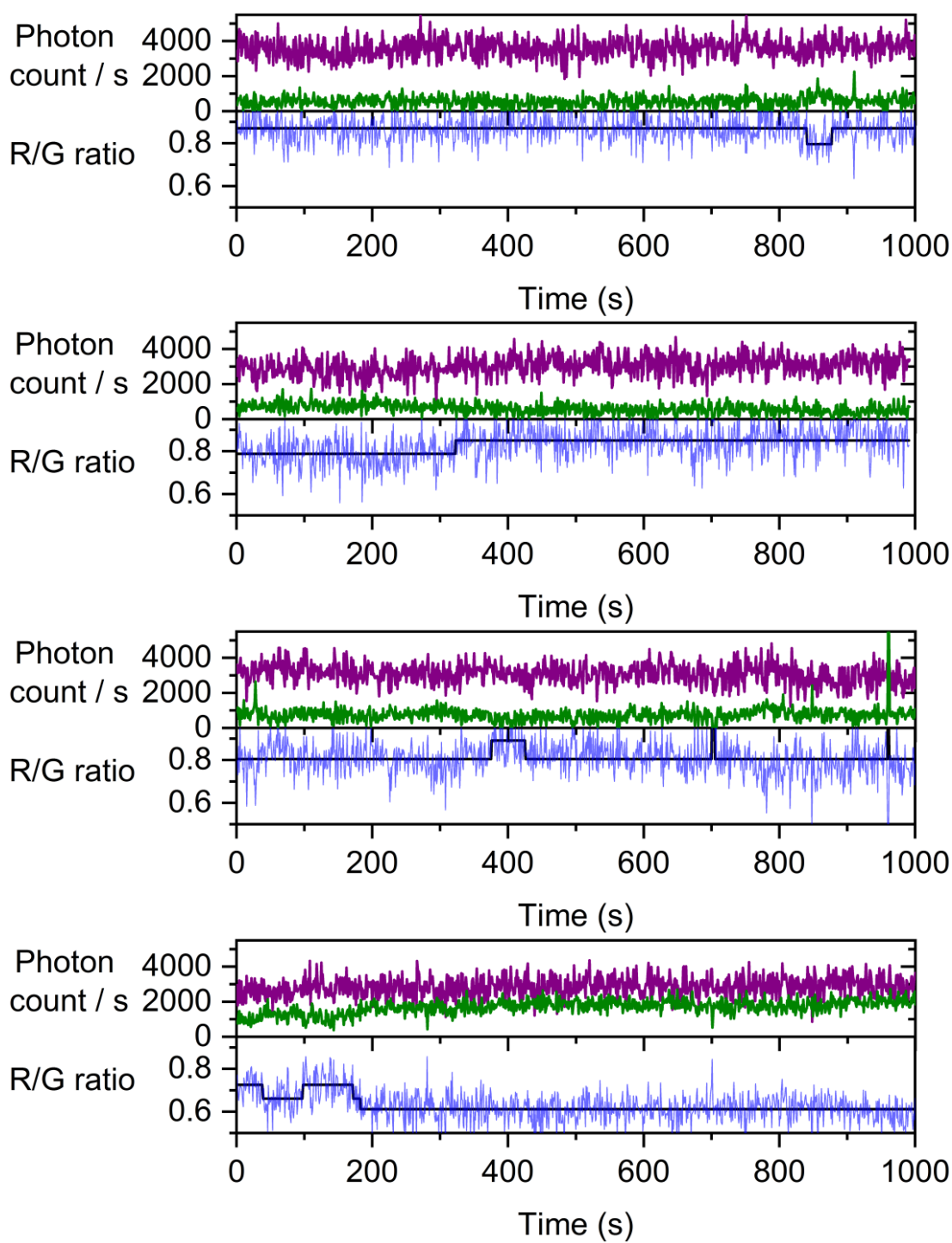

**Figure S7: Typical rare-fluctuation dynamic traces.** The immobilized NDs were imaged in air using 532 nm excitation 3.4 kW cm$^{-2}$ and 1 s exposure for 20 min to 60 min.

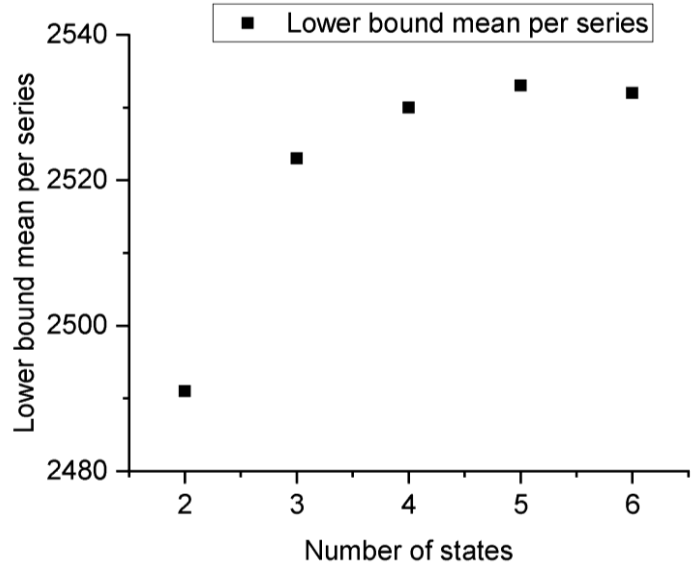

**Figure S8: plot of the mean values for lower bound per series for the models with different number of states as extracted from the HMM analysis.** As described by van de Meent *et al.*[34], the lower bound is used to compare the fitting of each model (*i.e.* one to 6 states). The higher the lower bound is and the better the fit is. As presented above, the 3-state model shows the largest increase in the lower bound (+30). Additional states led to only smaller improvements, for this reason and because we mainly observed three states in our analysis of static traces, we decided to use a 3-state model. The standard deviation is 1110 per point.

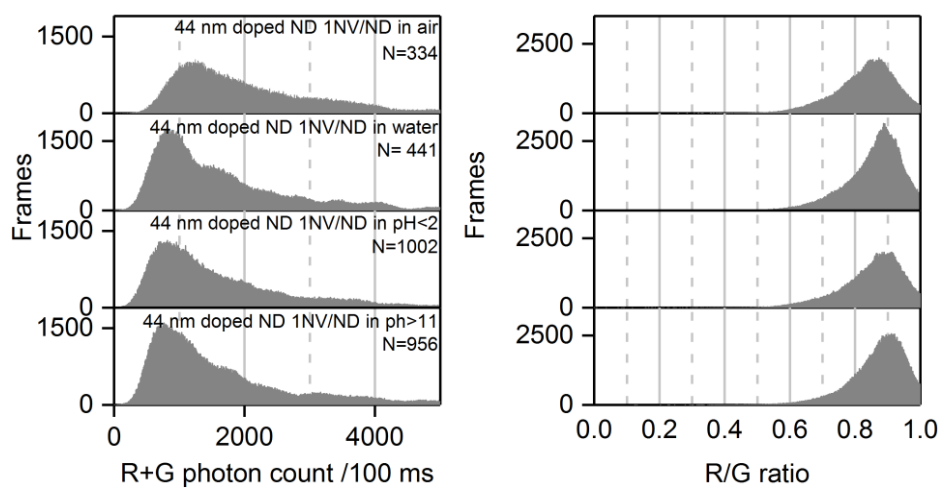

**Figure S9: distribution of photon count and R/G ratio from 44-nm doped NDs (1NV/ND) in different solutions**. The effect of pH in changing the charge state could not be reproduced in the 44-nm NDs. As shown in the figure above, the two bottom rows (in acid and base) show no significant change in the photon count distribution and R/G ratio as observed in the 10-nm NDs (see Figure 5). A 0.01 M HCl solution was used for pH <2 measurement and a 0.1 M NaOH for pH >11. N is the number of fluorescent NDs observed and "Frames" is the frequency of the measurement from a ND (see Figure 1 for details). The immobilized NDs were imaged with 532 nm excitation 7.8 kW cm$^{-2}$ 100 ms exposure for 25 s.

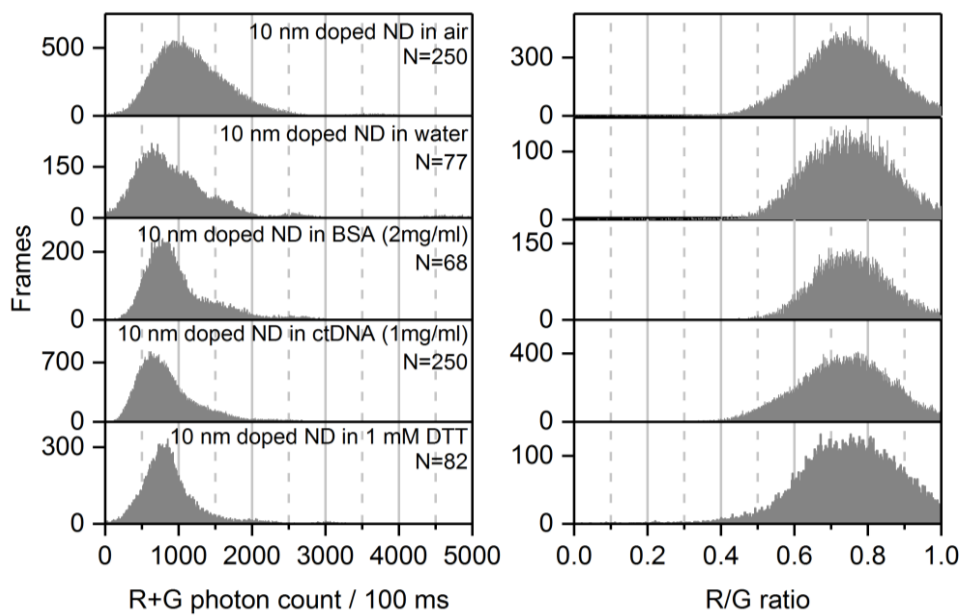

**Figure S10: distribution of photon count and R/G ratio from 10-nm doped NDs in different solutions.** No effect of different solutions (reducing agent DTT: Dithiothreitol, BSA: bovine serum albumin, and ctDNA: calf thymus DNA) on the charge state was found on the 10-nm NDs. As shown in the figure above, the three bottom rows (in DTT, BSA and DNA) show similar photon count distribution and R/G ratio compared to water (second row). N is the number of fluorescent NDs observed and "Frames" is the frequency of the measurement from a ND (see Figure 1 for details). The immobilized NDs were imaged with 532 nm excitation 7.8 kW cm$^{-2}$ and 100 ms exposure for 25 s.

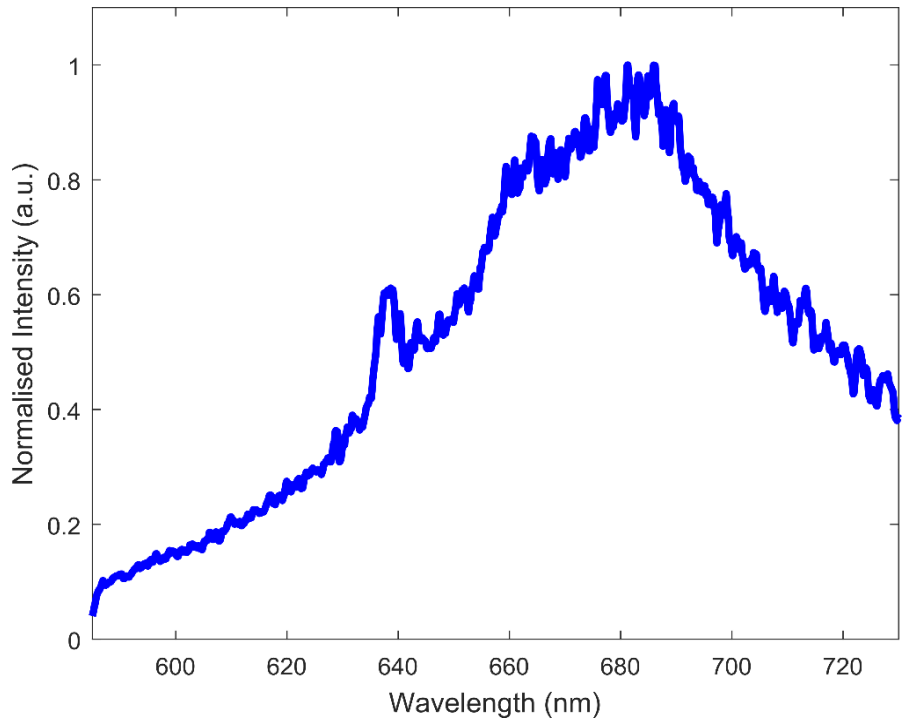

**Figure S11: emission spectrum of NV⁻ from the 44-nm doped NDs (1NV/ND) with 532 nm laser excitation.** The presence of NV center in NDs is verified by the manufacturer in the doped samples (10, 40, 44 and 200-nm). The emission spectra of four 44-nm doped NDs were also measured. As shown above, they all show typical NV⁻ emission spectra with the characteristic zero-phonon-line at 637 nm.